\documentclass[11pt]{article}

\usepackage[utf8]{inputenc}
\usepackage[T1]{fontenc}
\usepackage[margin=1in]{geometry}
\usepackage{amsmath,amssymb}
\usepackage{booktabs}
\usepackage{graphicx}
\usepackage{tikz}
\usetikzlibrary{positioning,arrows.meta}
\usepackage{listings}
\usepackage{xcolor}
\usepackage{microtype}
\usepackage[round,authoryear]{natbib}
\usepackage[hidelinks]{hyperref}
\usepackage{url}
\usepackage{enumitem}
\usepackage{authblk}
\usepackage{placeins}
\usepackage{float}  

\definecolor{codebg}{RGB}{246,247,249}
\title{\bf Token-Native Storage:\\
Read and Write in your Agent's Language}

\author[1]{Kumar Shivendu}
\affil[1]{Qdrant \\ \texttt{mail@kshivendu.dev}, \texttt{kumar.shivendu@qdrant.com}}

\date{\today}

\begin{document}
\maketitle

\begin{abstract}
Search and database engines still store text as UTF-8, a format built for humans. But the
systems that increasingly read and write that text (embedders, rerankers, and
language-model agents) work with \emph{token IDs}, not characters, so every access pays to
translate between the two. As agents become the primary readers and writers of stored text,
we argue for \emph{token-native storage}: keep the text as the model's own
byte-pair-encoding (BPE) token IDs. Packing r50k IDs as
\texttt{uint16} already beats UTF-8 by $2.25\times$ on English with no compression, and an
entropy coder on top reaches $3.30\times$. Across six tokenizers and three corpora (English, code,
Hindi), compressing token IDs matches or beats every byte codec, even a corpus-trained zstd
dictionary. Two findings sharpen the case. BPE numbers tokens by merge order instead of frequency,
and re-ranking by frequency lets a plain integer codec (streamvbyte) recover most of the
entropy coder's ratio while decoding ${\sim}7\times$ faster, a near-free change to how AI labs publish vocabularies. And because a model reads token IDs, not text, a
token-native store hands over the IDs directly instead of re-tokenizing on every read. The only requirement is that reader and writer share a tokenizer, and different model families often use different ones today, so we argue for 
standardization: a published, shared vocabulary, the way ASCII and UTF-8 standardized text.
\end{abstract}

\section{Introduction}

Search and database engines store text as UTF-8 bytes,\footnote{UTF-8 is the near-universal default. A store may compress those bytes, but the
representation stays UTF-8, whereas token-native storage changes the representation itself. All
ratios are over UTF-8.} a human-facing format. But a
retrieval or agentic system rarely reads those bytes as UTF-8: the moment a model
touches the text it converts the bytes to token IDs, and converts them back while writing.
The infrastructure therefore keeps the same content in two representations and pays for
both: storage for the UTF-8 bytes, and a tokenization pass on every read.


We argue for a different default for agents: \emph{token-native storage}. Store the text
as the serving model's own BPE token IDs, the representation every consumer
(embedder, reranker, LLM) already needs. Agents are becoming the primary readers
and writers of stored text, and will soon account for more of it than humans: retrieval-augmented
generation, chat history, and the model's own training/finetuning data. For
them, a UTF-8 store forces a translation on every access, while a token-native store lets
them read and write in token IDs directly, at higher compression and throughput
(Section~\ref{sec:latency}). Translation then happens only at the edges, where the system
meets a human.

Beyond removing that translation cost, \textbf{a tokenizer that covers the script
compresses for free.} A BPE token is a learned unit of language, so packing token IDs into
fixed-width integers beats UTF-8 before any compression algorithm runs
(Section~\ref{sec:eval}).

Two further contributions distinguish this work from the off-the-shelf compression it
builds on:
\begin{itemize}[leftmargin=1.4em,itemsep=2pt]
\item \textbf{BPE token IDs are not frequency-ordered.} BPE assigns IDs in
merge-discovery order rather than by usage. Re-ranking IDs by corpus frequency lets a
variable-length integer codec (streamvbyte) recover most of the entropy coder's ratio at
${\sim}7\times$ faster decode (Section~\ref{sec:freq}). This motivates a near-free change
to how AI labs publish tokenizers: emit IDs in frequency order (Section~\ref{sec:ecosystem}).
\item \textbf{The latency win is an agent-workload win.} When the reader or writer
is a model, tokenization is mandatory. Token-native storage serves the IDs that
already exist instead of re-tokenizing on every read (Section~\ref{sec:latency}).
\end{itemize}

Compressing token IDs with off-the-shelf coders is prior art, frequency ordering included~\citep{kalcher2026}, which raises the compression ratio a general-purpose compressor reaches. We optimize instead for a live database: \texttt{+freq+vbyte}
keeps only streamvbyte, so reads decode in microseconds, and we argue AI labs should fix the ordering before shipping, so every user gets more compression without re-deriving it. Our broadest contribution is the perspective itself: token IDs as the read/write form of a model-facing database, so the store and model pay no translation cost. The only
limitation is portability: token IDs are shared only when reader and writer use the same
tokenizer, and different model families often use different tokenizers today. We give a database interface for using
token-native storage today, and argue that portability across models needs standardization
(Section~\ref{sec:ecosystem}).

\section{Related Work}
\label{sec:related}
Token-native storage sits next to three lines of work.

\paragraph{Storing token IDs, but as a static training file.}
Storing token IDs to avoid recomputing them is already common in training. NVIDIA's
Megatron-Core stores a tokenized corpus as a memory-mapped \texttt{.bin} of token IDs
with a \texttt{.idx} of document offsets, so
training never re-tokenizes~\citep{megatron}. But those IDs are a write-once training
input, read only by the model. We make token IDs a first-class field type in a database or search engine, where individual items are inserted, read, updated, and deleted on
demand. Each item is served either as IDs to a model or detokenized to characters for a human.

\paragraph{Removing the two representations, but with lossy ones.}
A line of work attacks the same waste by making retrieval and
generation share one \emph{learned} representation: continuous memory
tokens~\citep{clara2025}, a single forward pass over both tasks~\citep{onegen2024},
or learned discrete codes~\citep{unisearch2025}. These are lossy and model-specific.
Token-native storage keeps the tokenizer's own IDs, which decode back to the exact
source text a store must return.

\paragraph{Compressing token IDs.}
Compressing token-ID streams is established: neural arithmetic
coding~\citep{lester2024}, zstd over packed BPE IDs for stored LLM
prompts~\citep{lopace2026}, and frequency-ordered IDs with a varint before a general
compressor~\citep{kalcher2026}. Yet a compressor is not needed to beat UTF-8: raw packed token IDs already compress on their own, \texttt{uint16} for r50k and three bytes for o200k. We benchmark
his best variant (zstd-22, since LZMA costs far more for a smaller gain): our entropy coder
edges it on natural language, our dictionary on code, and \texttt{+freq+vbyte} keeps most of its
ratio with no general-purpose compressor, just streamvbyte, and decodes about $7\times$ faster. Across all three,
tokenization is considered a preprocessing step for compression.

\paragraph{Positioning.}
None of these proposes token IDs as the canonical and stored form of text for models,
with on demand and fast read/write access to individual items, an exactly-invertible
vocabulary, all while maintaining higher compression than utf-8.

\section{Token-Native Storage}
\label{sec:design}

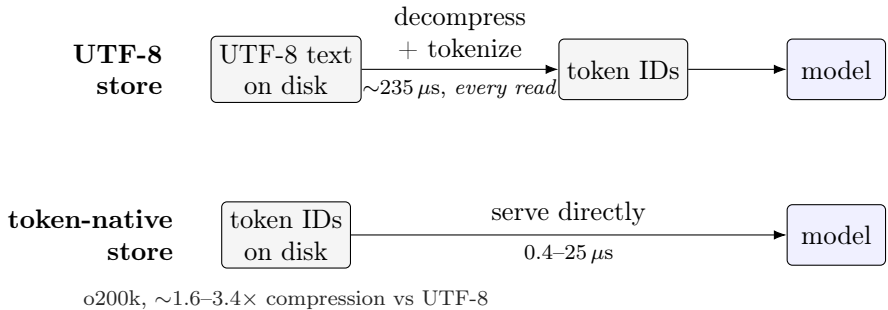
\begin{figure}[ht]
\centering
\begin{tikzpicture}[
  font=\small, >=Latex, node distance=0mm,
  disk/.style={draw, rounded corners=2pt, minimum height=8mm, minimum width=17mm,
    align=center, inner sep=3pt, fill=gray!8},
  model/.style={draw, rounded corners=2pt, minimum height=8mm, minimum width=13mm,
    align=center, inner sep=3pt, fill=blue!6},
]
\node[disk] (bdisk) {UTF-8 text\\on disk};
\node[disk, right=26mm of bdisk] (bids) {token IDs};
\node[model, right=13mm of bids] (bmodel) {model};
\draw[->] (bdisk) -- node[above, align=center]{decompress\\+ tokenize}
  node[below, align=center]{\scriptsize ${\sim}235\,\mu$s, \emph{every read}} (bids);
\draw[->] (bids) -- (bmodel);
\node[left=5mm of bdisk, align=right, text=black] {\bfseries UTF-8\\\bfseries store};
\node[disk, below=13mm of bdisk] (tdisk) {token IDs\\on disk};
\node[model] at (bmodel |- tdisk) (tmodel) {model};
\draw[->] (tdisk) -- node[above, align=center]{serve directly}
  node[below, align=center]{\scriptsize $0.4$--$25\,\mu$s} (tmodel);
\node[left=5mm of tdisk, align=right, text=black] {\bfseries token-native\\\bfseries store};
\node[below=1.5mm of tdisk, text=black!85, align=center] {\scriptsize o200k, ${\sim}1.6$--$3.4\times$ compression vs UTF-8};
\end{tikzpicture}
\caption{Every model read from a UTF-8 store pays a decompress-and-tokenize cost
(${\sim}235\,\mu$s), while a token-native store serves the token IDs directly
($0.4$--$25\,\mu$s). Writes mirror this:
a model emits token IDs, which a UTF-8 store must first detokenize and compress.}
\label{fig:schematic}
\end{figure}

\subsection{Step 1: store the token IDs}
On write, tokenize the text with the serving model's BPE tokenizer~\citep{sennrich2016} and store the
integer token-ID sequence. On read, return the IDs to a token-consuming client or
detokenize for a human. When the writer is itself a model, the IDs already
exist (the sampler produced them).

The tokenizer is a collection-level choice: a shared vocabulary and frequency table
serve one tokenizer at a time. Text may be read by models with different tokenizers
(an embedder, a reranker, an LLM), so store the IDs in the one that reads and writes
most, usually the serving LLM. A consumer with a different tokenizer detokenizes and
re-tokenizes, one detokenization more than reading UTF-8. That detokenization runs in tens of
microseconds, well below the re-tokenization both paths pay. Full portability across models will come with a shared vocabulary (Section~\ref{sec:ecosystem}).

Tokenization compresses before any algorithm runs. Here's a way to understand it. A BPE token covers about
three-quarters of an English word on average (a common rule of thumb) and a word
is about six UTF-8 bytes (5-character average + 1 space), so
\[
\underbrace{6\,\text{B/word}}_{\text{text}}\times
\underbrace{\tfrac34\,\text{word/token}}_{\text{BPE}}=4.5\,\text{B/token}
\;\text{vs.}\;\underbrace{2\,\text{B/token}}_{\text{uint16 ID}}
\Rightarrow{\sim}2.25\times .
\]
We use OpenAI's \texttt{tiktoken} tokenizers~\citep{tiktoken}: r50k
($50{,}257$ IDs, \texttt{uint16}), cl100k ($100{,}277$) and o200k ($200{,}019$).
The larger two exceed \texttt{uint16}. The obvious fallback, \texttt{uint32}, wastes a
byte: four bytes per ID barely beats UTF-8's $4.5$. Three-byte packing keeps the win and
still covers any vocabulary, since $24$ bits hold $16.7$M IDs (Gemma's $256{,}000$ tokens
use only $18$). This 3-byte packing, not \texttt{uint32}, is the practical raw representation.

\subsection{Step 2 (optional): compress the token IDs}
\label{sec:coder}
A byte codec cannot know that \texttt{0x017F} is token $383$. Several optional,
interchangeable coders instead run on the packed IDs, all lossless and all decoding
straight back to token IDs with no re-tokenization. They trade ratio against decode speed:
\begin{itemize}[leftmargin=1.4em,itemsep=2pt]
\item \textbf{\texttt{+freq+vbyte}} (recommended default): frequency-remap the IDs, then pack with
streamvbyte, a SIMD varint. Most of ANS's ratio at a fraction of the decode cost, with no
general-purpose compressor (Section~\ref{sec:freq}).
\item \textbf{\texttt{+ans}}: a static unigram entropy coder~\citep{duda2013} (via
\texttt{constriction}, \citealp{bamler2022}) over the IDs, its table trained once and shared
across documents. Best ratio on natural language, slower decode.
\item \textbf{\texttt{+zdict}} and \textbf{\texttt{+lz4}}: a corpus-trained zstd dictionary or
LZ4 over the packed IDs, the token-domain parallels of the byte \texttt{zstd --train} and
\texttt{LZ4} codecs. \texttt{+zdict} wins on repetitive data such as code, but ships its
dictionary with the data.
\end{itemize}
Raw packing already compresses on its own (Section~\ref{sec:eval}), so a coder is optional.
What sets all of these apart on read: they decode straight to token IDs, whereas a byte codec
must re-tokenize the text it decompresses.

\section{The frequency-ordering fix}
\label{sec:freq}
BPE numbers its vocabulary inefficiently. Each new token gets the next integer ID as it
is discovered during merging, so IDs reflect merge order instead of usage: a common token can
sit at ID $40{,}000$ while a rare one sits
at ID $400$. Variable-length integer codecs (streamvbyte, LEB128) spend fewer bytes on
small integers, so this numbering leaves compression on the table for free.
\citet{kalcher2026} also reorders BPE IDs by frequency, but only as a preprocessing
step for a general-purpose compressor. We instead store the reordered IDs and decode them with
streamvbyte alone, skipping the general-purpose compressor for a faster decode.

The \texttt{+freq+vbyte} method re-ranks IDs by corpus frequency (most frequent token gets
the smallest ID), then packs with \emph{streamvbyte}, a SIMD varint codec. On English (C4), it
reaches $2.73\times$ (median, o200k), between raw packing ($1.59\times$) and ANS
($3.40\times$), while decoding ${\sim}7\times$ faster than ANS
(Section~\ref{sec:latency}). The re-rank is what does the work: streamvbyte spends fewer
bytes on small integers, so it pays off only once the frequent tokens carry the smallest
IDs. On the original merge-order IDs, which are large, it barely improves on fixed-width packing.

The fix generalizes beyond our codec: any downstream integer compressor benefits, at
no modeling cost. This motivates a one-line request to tokenizer authors: publish
vocabularies with IDs in frequency order (Section~\ref{sec:ecosystem}). Absent that,
users can re-rank on their own corpus and beat the vendor's assigned token IDs.

\section{Evaluation}
\label{sec:eval}
\begin{figure}[ht]
\centering
\includegraphics[width=\linewidth]{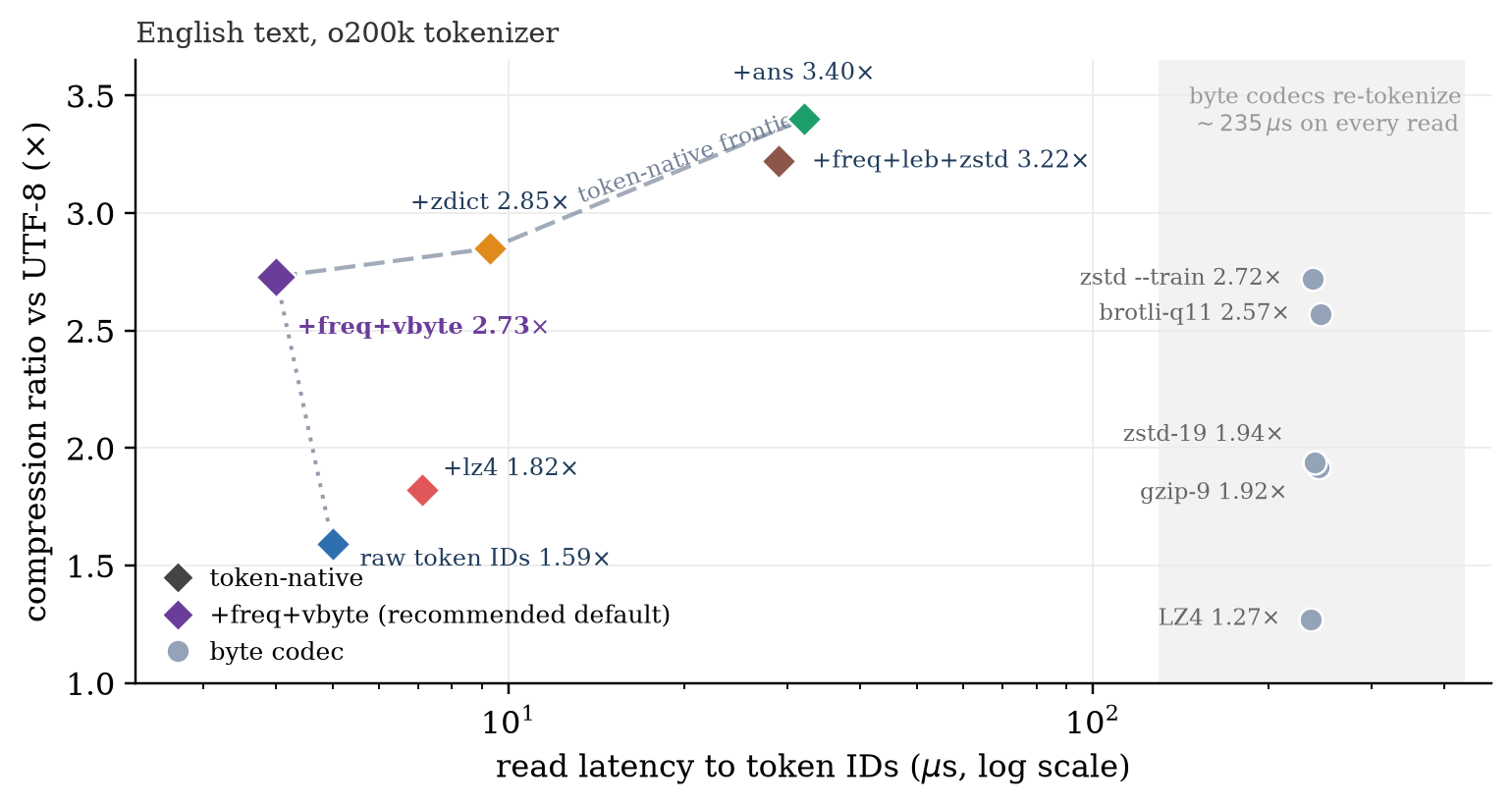}
\caption{Token-native methods form the ratio-latency frontier (English, o200k, the
tokenizer GPT-4o-class models use). Read latency to token IDs (log $x$) against
compression ratio, with a line through the token-native Pareto frontier. Every byte
codec sits near ${\sim}235\,\mu$s because a model must re-tokenize the text on read,
while token-native methods return IDs in microseconds. \texttt{+freq+vbyte} is the fast,
high-ratio recommended default.}
\label{fig:frontier}
\end{figure}

We evaluate on three real, held-out, non-overlapping corpora: English
(C4)~\citep{c4}\footnote{\url{https://huggingface.co/datasets/allenai/c4}},
Python code (codeparrot-clean)\footnote{\url{https://huggingface.co/datasets/codeparrot/codeparrot-clean}},
and Hindi (Wikipedia)\footnote{\texttt{wikimedia/wikipedia}, config
\texttt{20231101.hi}: \url{https://huggingface.co/datasets/wikimedia/wikipedia}}. We use $512$-token chunks (a standard embedding chunk size). Each domain has an $8$M-token train split for the static tables and a $1.5$M-token held-out test split. Training (ANS table, zstd dictionary, frequency ranks) uses
only each domain's train split.\footnote{Held-out test ratios stay within ${\sim}2$--$9\%$ of train, so the tables do not overfit, and hold within ${\sim}2\%$ when scaled to ${\sim}1$\,GB per domain.} Every method is lossless. Code and benchmarks are
available at \url{https://github.com/kshivendu/token-storage}.
Figure~\ref{fig:frontier} previews where each method lands on compression ratio and
read latency.

\subsection{Compression ratio}
A tokenizer that covers the script compresses for free (Table~\ref{tab:ratio}). Raw
token-ID packing reaches $2.25\times$ on English and $2.55\times$ on Hindi, ahead of LZ4,
gzip, and zstd, with no algorithm running. Used off its script a tokenizer does poorly
and can even expand (r50k raw is $0.84\times$ on Hindi), which an entropy coder or a
script-appropriate tokenizer repairs.

\begin{figure}[ht]
\centering
\includegraphics[width=0.8\linewidth]{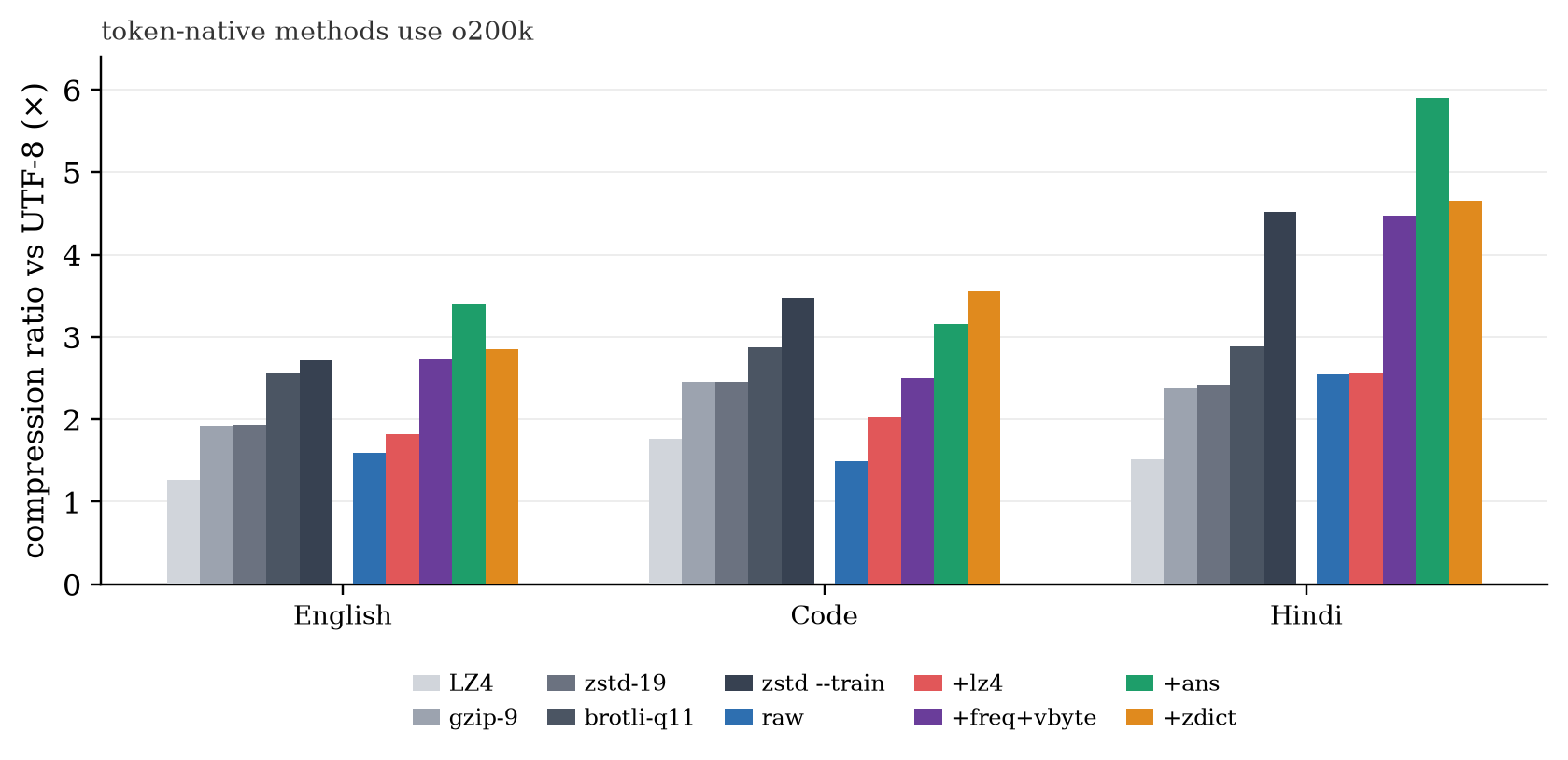}
\caption{Median compression ratio over UTF-8 by method, one bar per corpus.
Token-native methods (colored) use o200k. Per-tokenizer numbers are in
Table~\ref{tab:ratio}.}
\label{fig:compression}
\end{figure}

Compressing the token IDs then matches or beats every byte codec (Figure~\ref{fig:compression}).
On natural language it wins outright: a static entropy coder over the IDs (\texttt{+ans})
reaches $3.30\times$ on English and $5.90\times$ on Hindi, past the strongest byte codec
\texttt{zstd --train}. Code is the exception, because its heavy repetition is something an
order-0 coder cannot exploit. There a corpus-trained dictionary over the packed IDs
(\texttt{+zdict}, $3.55\times$) wins instead, though such dictionaries are corpus-specific and
ship with the data, unlike the shared tokenizer. The coders trace a ratio/decode-speed
frontier (Figure~\ref{fig:frontier}) that token-native storage can pick any point on.

\begin{figure}[ht]
\centering
\includegraphics[width=0.56\linewidth]{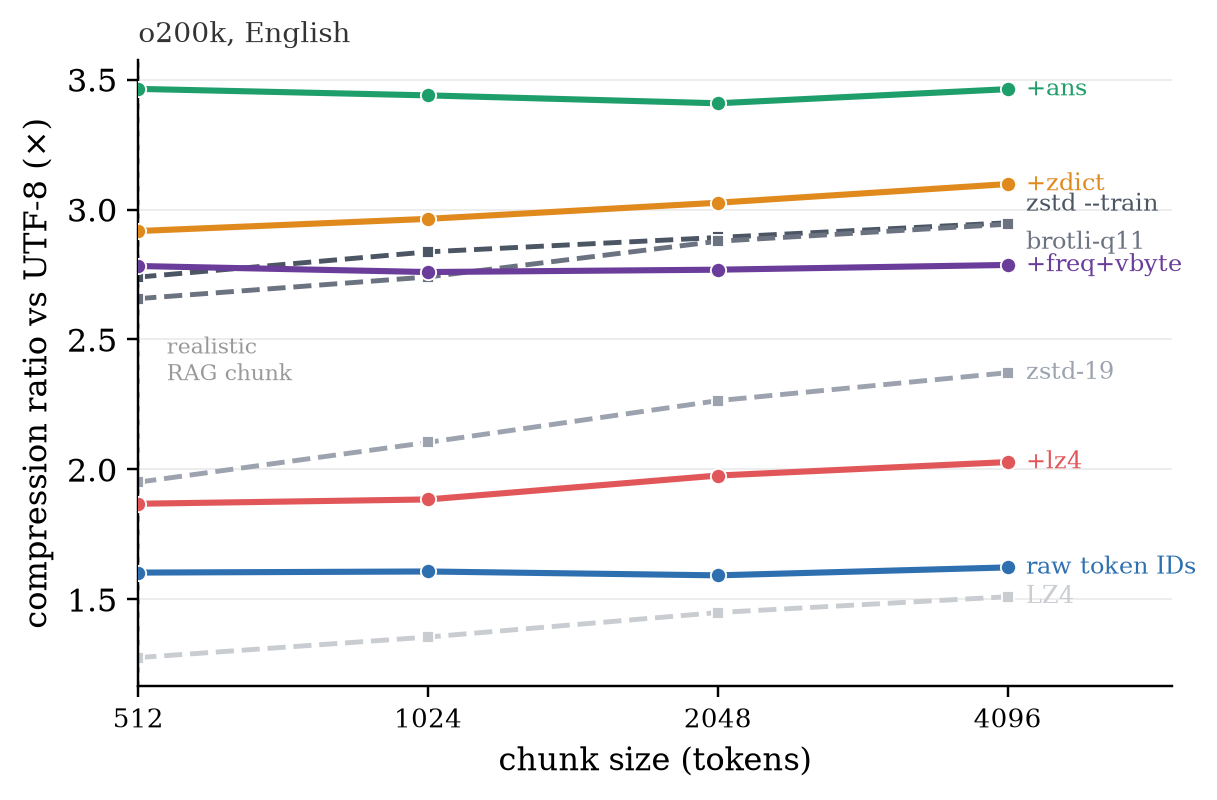}
\caption{Compression ratio vs.\ chunk size (o200k, English). Order-0 token methods
(\texttt{raw}, \texttt{+ans}, \texttt{+freq+vbyte}) stay flat because per-token entropy is
additive. LZ-family methods (byte codecs, \texttt{+zdict}, \texttt{+lz4}) climb by exploiting
cross-chunk redundancy: a corpus-trained \texttt{zstd --train} catches \texttt{+freq+vbyte} only at
$4096$ tokens, while \texttt{+ans} and \texttt{+zdict} stay on top at every size. The token-native
advantage is largest at the realistic $512$-token chunk.}
\label{fig:scaling}
\end{figure}

\texttt{+freq+vbyte} gives most of ANS's ratio at the fastest decode: frequency-ordered IDs
packed with streamvbyte and \emph{no} general-purpose compressor. Order-0 ratios barely move with
chunk size, so the token-native ratio advantage is clearest at realistic $512$-token chunks
(Figure~\ref{fig:scaling}) while its latency advantage holds at every size. The full grid
across tokenizers, corpora, and sizes is in the repository.

\begin{table}[ht]
\centering
\caption{Median compression ratio vs.\ raw UTF-8, $512$-token chunks, static tables
trained on each corpus's train split and measured on held-out test
(C4 English / codeparrot / Hindi Wikipedia). Byte codecs (top) are
tokenizer-independent. Token-native methods (bottom) are shown per tokenizer.
\texttt{+freq+leb+zstd} is Kalcher's method~\citep{kalcher2026}: frequency-ordered IDs, a LEB128 varint, then a general-purpose compressor.\texttt{+lz4} and \texttt{+zdict} apply LZ4 and a corpus-trained zstd dictionary to the
packed token IDs, the token-domain parallels of the byte \texttt{LZ4} and
\texttt{zstd --train} rows, evaluated on held-out test. Bold marks the best per language: \texttt{+ans} on English and Hindi,
\texttt{+zdict} on code, where cross-file repetition lets the dictionary help.}
\label{tab:ratio}
\small\setlength{\tabcolsep}{4.5pt}
\begin{tabular}{l|ccc|ccc|ccc}
\toprule
Method & \multicolumn{3}{c|}{English} & \multicolumn{3}{c|}{Code} & \multicolumn{3}{c}{Hindi} \\
\midrule
LZ4 & \multicolumn{3}{c|}{$1.27$} & \multicolumn{3}{c|}{$1.76$} & \multicolumn{3}{c}{$1.52$} \\
gzip \texttt{-9} & \multicolumn{3}{c|}{$1.92$} & \multicolumn{3}{c|}{$2.46$} & \multicolumn{3}{c}{$2.38$} \\
zstd \texttt{-19} & \multicolumn{3}{c|}{$1.94$} & \multicolumn{3}{c|}{$2.45$} & \multicolumn{3}{c}{$2.42$} \\
brotli \texttt{q11} & \multicolumn{3}{c|}{$2.57$} & \multicolumn{3}{c|}{$2.87$} & \multicolumn{3}{c}{$2.89$} \\
zstd \texttt{--train} & \multicolumn{3}{c|}{$2.72$} & \multicolumn{3}{c|}{$3.47$} & \multicolumn{3}{c}{$4.52$} \\
\midrule
& r50k & cl100k & \textbf{o200k} & r50k & cl100k & \textbf{o200k} & r50k & cl100k & \textbf{o200k} \\
\cmidrule(lr){2-4}\cmidrule(lr){5-7}\cmidrule(lr){8-10}
\textbf{raw} & $2.25$ & $1.55$ & $1.59$ & $1.07$ & $1.49$ & $1.49$ & $0.84$ & $0.86$ & $2.55$ \\
\textbf{\texttt{+freq+vbyte}} & $2.60$ & $2.69$ & $2.73$ & $1.41$ & $2.51$ & $2.50$ & $1.33$ & $2.04$ & $4.47$ \\
\texttt{+lz4} & $2.34$ & $1.80$ & $1.82$ & $2.10$ & $2.03$ & $2.02$ & $1.34$ & $1.32$ & $2.57$ \\
\texttt{+ans} & $3.30$ & $3.37$ & $\mathbf{3.40}$ & $2.58$ & $3.19$ & $3.16$ & $2.97$ & $3.48$ & $\mathbf{5.90}$ \\
\texttt{+freq+leb+zstd} &$3.09$ & $3.19$ & $3.22$ & $3.01$ & $3.35$ & $3.35$ & $2.71$ & $3.04$ & $4.80$ \\
\texttt{+zdict} & $3.15$ & $2.85$ & $2.85$ & $3.27$ & $3.36$ & $\mathbf{3.55}$ & $4.33$ & $4.31$ & $4.65$ \\
\bottomrule
\end{tabular}
\end{table}

\subsection{Generality across tokenizers}
The effect is not an OpenAI or vocab-size artifact. On English, six modern tokenizers
(r50k, cl100k, o200k, Qwen2.5, DeepSeek-V2, Gemma-2) land in a tight $3.30$--$3.40\times$
band with static ANS. Raw packing is fixed-width, so it tracks vocabulary size instead:
$2.27\times$ for r50k's 2-byte IDs and $1.50$--$1.60\times$ for the five 3-byte
vocabularies. Compression flattens that spread. Latency is stable across tokenizers too: \texttt{+freq+vbyte} reads in ${\sim}4\,\mu$s and \texttt{+ans} in
${\sim}30\,\mu$s, with only raw packing sensitive to vocabulary (\texttt{uint16} reinterpret
for r50k, 3-byte unpack for the larger two).

\subsection{The tokenizer drives most of the gain}
Most of the gain is the \emph{tokenizer}, not the coder. Feeding a coder token IDs instead
of UTF-8 bytes is what lifts the ratio: ANS over raw UTF-8 bytes reaches only $1.76\times$
on English, but over tokens it jumps to $3.30\times$. LZ-family coders (zstd, \texttt{+lz4})
gain the same way, though less, because BPE already collapses each frequent character
sequence into one short token, leaving fewer long repeats in the stream for their
substring matching to find.

\FloatBarrier
\subsection{Latency: an agent-workload win}
\label{sec:latency}

\begin{figure}[ht]
\centering
\includegraphics[width=\linewidth]{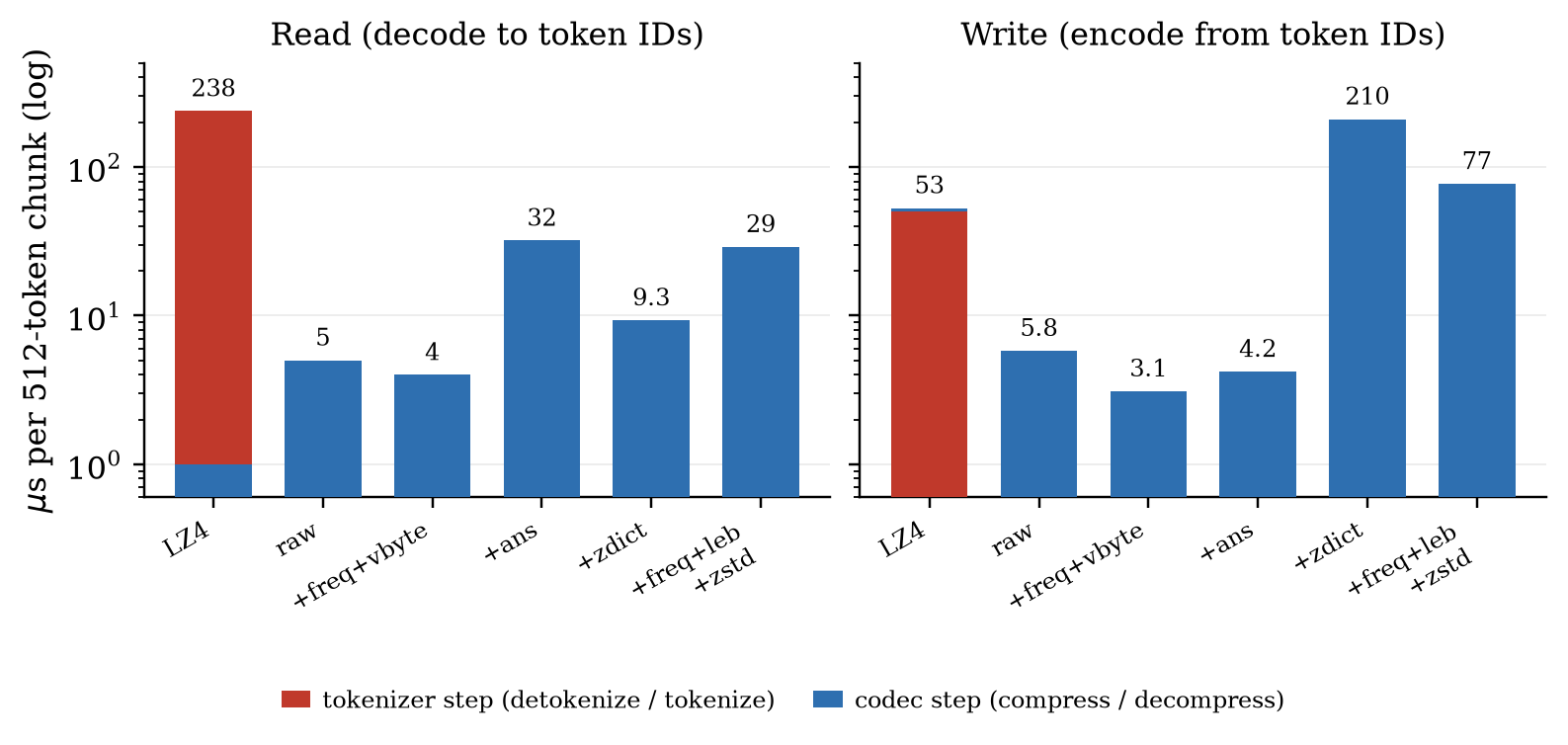}
\caption{Where read/write latency goes, decomposed at the o200k basis (English, median
per $512$-token chunk, \textbf{agent} reader/writer). A byte codec (LZ4) pays a tokenizer step
on every access: $237\,\mu$s to tokenize on read and $50\,\mu$s to detokenize on write. Token-native
methods skip that step, so their bars are only the codec step: microseconds for raw,
\texttt{+freq+vbyte}, and \texttt{+ans}, more for the compressor-backed \texttt{+zdict} and
\texttt{+freq+leb+zstd} (Kalcher).}
\label{fig:latency}
\end{figure}

Tokenization is expensive (${\sim}235\,\mu$s per fresh $512$-token chunk with
\texttt{tiktoken} o200k on a single core, against ${\sim}8\,\mu$s for LZ4 compress).\footnote{Timed on distinct chunks with the tokenizer. The cost varies little across tokenizers: within ${\sim}1.2\times$ for
r50k, cl100k, and o200k. Detokenization is $2$--$6\times$ cheaper.} When the primary reader is a model, \textbf{tokenization is mandatory} on every
read and no faster tokenizer drives it to zero, so the only question is whether you pay it
every time or once. 

\begin{itemize}[leftmargin=1.4em,itemsep=2pt,topsep=3pt]
\item \emph{Write}: an LLM emits token IDs already, so token-native storage packs them in
microseconds, whereas any utf-8 store must first detokenize and then compress.
\item \emph{Read}: token IDs are returned directly in microseconds, while UTF-8 text must
be decompressed and tokenized back for the model (${\sim}235\,\mu$s per chunk).
\end{itemize}
Example: A search query returns multiple records: at $100$ chunks per query, with a
UTF-8 store, the system spends ${\sim}24$\,ms of serial CPU re-tokenizing text for the model.

\begin{table}[ht]
\centering
\caption{Median \textbf{agent} write / read latency per $512$-token chunk (English, $\mu$s).
Byte codecs (top) are split into their steps at the o200k basis: write detokenizes then
compresses, read decompresses then re-tokenizes (${\sim}237\,\mu$s), so the bold \textbf{total}
is dominated by the tokenizer step. Token-native methods (bottom) skip that step and serve
IDs directly, staying in the $\mu$s range across all three tokenizers. Bold marks the fastest
token-native write and read.}
\label{tab:latency}
\vspace{4pt}
\small\setlength{\tabcolsep}{3.5pt}
\begin{tabular}{l|ccc|ccc}
\toprule
Method & \multicolumn{3}{c|}{write ($\mu$s)} & \multicolumn{3}{c}{read ($\mu$s)} \\
\cmidrule(lr){2-4}\cmidrule(lr){5-7}
 & detok. & compress & \textbf{total} & decomp. & tokenize & \textbf{total} \\
\midrule
LZ4 & $50$ & $2.9$ & $53$ & $1.0$ & $237$ & $238$ \\
gzip-9 & $50$ & $26$ & $76$ & $8.0$ & $237$ & $245$ \\
zstd-19 & $50$ & $209$ & $260$ & $4.5$ & $237$ & $241$ \\
zstd \texttt{--train} & $50$ & $360$ & $410$ & $3.2$ & $237$ & $240$ \\
brotli-q11 & $50$ & $2777$ & $2828$ & $9.5$ & $237$ & $246$ \\
\midrule
 & r50k & cl100k & \textbf{o200k} & r50k & cl100k & \textbf{o200k} \\
\cmidrule(lr){2-4}\cmidrule(lr){5-7}
\textbf{raw} & $\mathbf{0.5}$ & $6.0$ & $5.8$ & $\mathbf{0.4}$ & $5.3$ & $5.0$ \\
\textbf{\texttt{+freq+vbyte}} & $2.9$ & $3.1$ & $3.1$ & $3.6$ & $4.1$ & $4.0$ \\
\texttt{+lz4} & $2.6$ & $9.5$ & $9.4$ & $2.1$ & $7.1$ & $7.1$ \\
\texttt{+ans} & $3.8$ & $4.2$ & $4.2$ & $25$ & $30$ & $32$ \\
\texttt{+zdict} & $77$ & $214$ & $210$ & $4.2$ & $9.5$ & $9.3$ \\
\texttt{+freq+leb+zstd} &$73$ & $83$ & $77$ & $27$ & $31$ & $29$ \\
\bottomrule
\end{tabular}
\end{table}

\texttt{+freq+vbyte} is the fast and recommended default: it decodes ${\sim}7\times$ faster than
\texttt{+ans} for about $20\%$ less compression on English. \texttt{+zdict} pays more on write, but it's good with compression for repetitive corpora such as code.

\begin{figure}[H]
\centering
\includegraphics[width=0.62\linewidth]{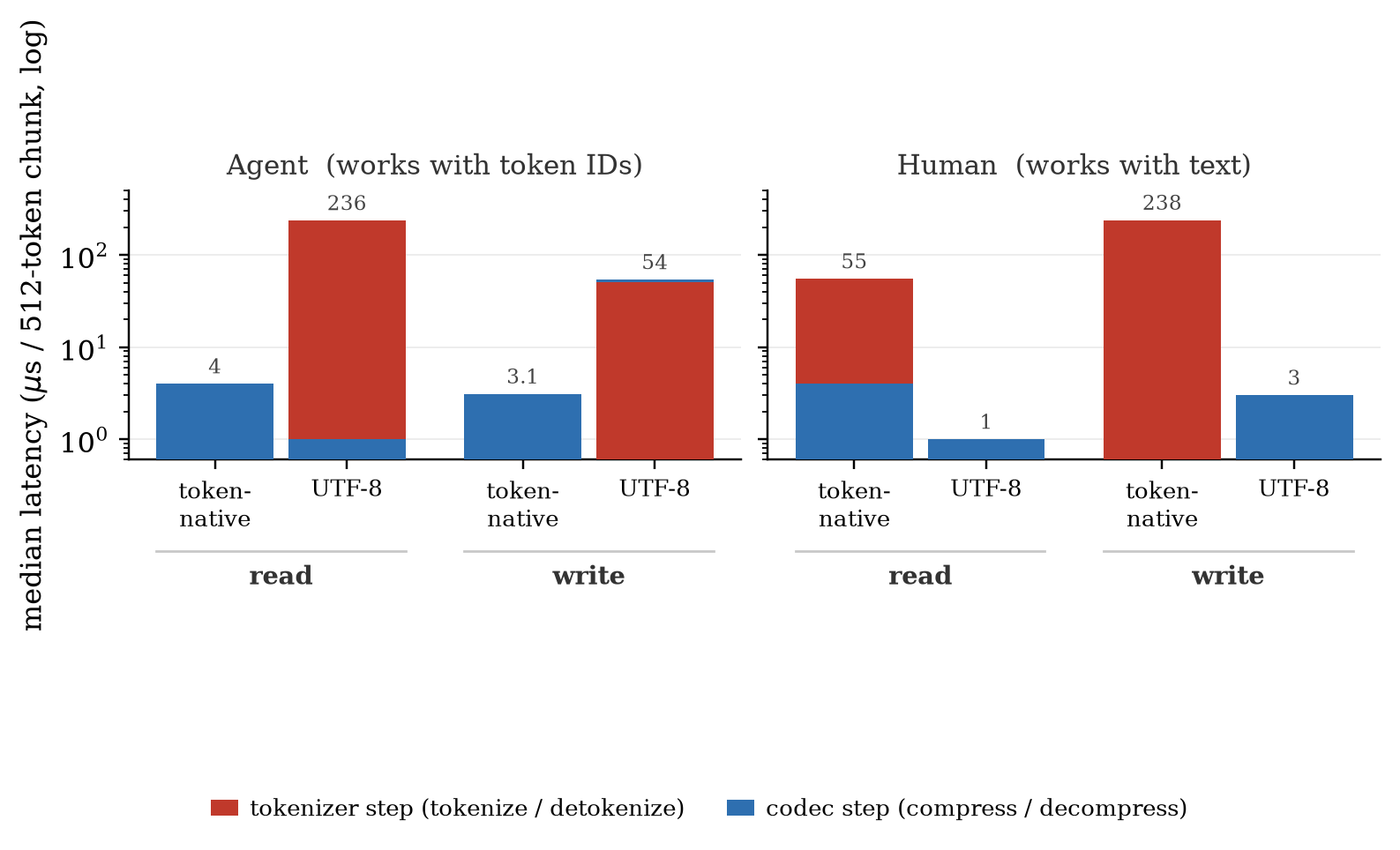}
\caption{The tokenizer step (red) is the cost that moves. For an agent it sits on the
UTF-8 store, which must tokenize on read and detokenize on write. For a human it flips onto
the token-native store. The codec step (blue) stays in the microseconds either way, so each
store is cheap for the party native to its stored form (o200k, \texttt{+freq+vbyte} on the
agent path).}
\label{fig:humanagent}
\end{figure}

An agent works with token IDs, so token-native storage is fast for it
and a UTF-8 store slow. But a human works with text, so the costs flip: a token-native store
tokenizes what a human writes (${\sim}235\,\mu$s) and detokenizes when a human
reads (${\sim}51\,\mu$s).

Token-native storage is therefore the right default when agents dominate the reads and
writes, the common case for agentic systems (Figure~\ref{fig:humanagent}). Agents produce
most of the stored text while humans type short instructions, and one agentic search or
RAG answer reads hundreds of chunks before responding.


The saving can compound across the pipeline. A retrieval stack tokenizes the same text
repeatedly: once to embed a document at ingest, again to rerank the candidates, and again for every agent that reads a retrieved chunk. When the
database, the embedding model, the reranker, and the agents all speak token IDs, the
system tokenizes once as a human submits text and detokenizes once as a human reads the
answer. Every internal hop passes IDs through unchanged. The system pays the tokenize/detokenize
cost only at its human/frontend interface.

\subsection{Downstream effects of using Token IDs instead of UTF-8 characters}
Text is one of the heaviest fields in a record: thousands of characters each consuming a byte,
while other fields take only a few bytes. That is why using token IDs instead of characters pays off even more. 
Every copy of that text shrinks by the same factor (snapshots, backups,
write-ahead logs, replicas, network), so tokenization also makes everything faster and cheaper.  
At $1$\,B documents ($1000$-word average, each having 5 chars average + 1 space), the raw text is $6.0$\,TB. A UTF-8 store using LZ4, the default in Qdrant and Elasticsearch (Postgres and others use similar LZ-family codecs), makes it $4.7$\,TB (${\sim}\$4.5$k/yr on SSD\footnote{At ${\sim}\$0.08$ per GB-month for general-purpose SSD (AWS EBS gp3). Rates from Napkin Math, \url{https://github.com/sirupsen/napkin-math}.}). Token-native storage cuts that to $3.8$\,TB raw (${\sim}\$3.6$k/yr) or $2.2$\,TB with o200k+freq+vbyte (${\sim}\$2.1$k/yr), roughly half the LZ4 footprint. This also makes I/O-bound work $2\times$ faster because less data has to be moved.


\section{Interface}
\label{sec:interface}
Adoption needs only a one-time, collection-level tokenizer choice:
\begin{lstlisting}[language=Java]
PUT /collections/documents/index
{ "schema": { "text": { "type": "token", "tokenizer": "o200k" } } }
\end{lstlisting}
Inserts accept token IDs (could be directly from a model) or a plain string (the engine tokenizes):
\begin{lstlisting}[language=Java]
PUT /collections/documents/points
{ "points": [{ "id": 123, "vector": [...],
    "payload": { "text": [1858, 6427, 20272, ...] } }] }   // or a string
\end{lstlisting}
Reads ask per field for tokens or text. Default \texttt{"text"} keeps existing
clients unchanged. An LLM pipeline asks for \texttt{"tokens"} and skips
detokenization:
\begin{lstlisting}[language=Java]
POST /collections/documents/points/search
{ "vector": [...], "limit": 10, "with_payload": { "text": "tokens" } }
// Response:
{"results": [{"id": 1, "text": [1858, 6427, 20272, 318, 257, 6997, ...]}]}
\end{lstlisting}

\section{What the ecosystem needs}
\label{sec:ecosystem}
Token-native storage is production-ready within a single-model stack, but its full
potential can be unlocked with three ecosystem changes.
\begin{enumerate}[leftmargin=1.4em,itemsep=2pt]
\item \textbf{A standardized, published tokenizer vocabulary} (an ASCII/UTF-8 for
tokens), so IDs are portable across models and versions rather than a
per-vendor, per-version lock-in. Most families open-source their tokenizers
(Anthropic and Google's proprietary models are the notable exceptions). Embedding
models have started converging on LLM tokenizers too, for example OpenAI's
\texttt{text-embedding-3} reuses cl100k and Jina's embeddings-v4 moved to a BPE
tokenizer.
\item \textbf{Token-ID-native inference APIs} that accept and return token IDs
directly, so a store hands IDs to the model and takes back the IDs it emits with no
detokenize-then-retokenize round trip. Open source stacks already do this: vLLM accepts
pre-tokenized \texttt{prompt\_token\_ids} and can return token IDs on its
OpenAI-compatible endpoints.\footnote{Token-ID input (\texttt{TokensPrompt}):
\url{https://docs.vllm.ai/en/stable/api/vllm/inputs/}. Returning token IDs to avoid
retokenization drift: \url{https://vllm.ai/blog/2025-10-22-agent-lightning}.} The
closed source AI labs should do the same.
\item \textbf{Publish IDs in frequency order} (Section~\ref{sec:freq}). It costs the AI labs
almost nothing to re-rank their future vocabulary by frequency on training or generated data, and it gives
every token ID compression algorithm more compression, without any extra latency/compute.
\end{enumerate}

\section{Limitations}
\label{sec:limitations}
\begin{itemize}[leftmargin=1.4em,itemsep=3pt]
\item \textbf{Shared vocabulary.} The IDs are portable only for consumers using the same
tokenizer, which differs across model families and vendors. The fix is standardization just like UTF-8 and ASCII standardized characters (Section~\ref{sec:ecosystem}).
\item \textbf{Tokenizer coverage.} A tokenizer with poor coverage of a script splits
it into many short tokens, so its raw IDs can exceed raw UTF-8 (r50k on Hindi,
$0.84\times$). UTF-8 itself spends multiple bytes per character on such scripts. A
tokenizer that covers the script turns that inefficiency into a larger win (o200k,
$2.55\times$ raw on Hindi). Even with a mismatched tokenizer, token ID compression
can perform better than UTF-8 (r50k on Hindi, $1.33\times$ +freq+vbyte and $2.97\times$ +ans).
\item \textbf{Corpus-specific tables.} The ANS table and frequency ranks can degrade
under high domain mismatch, so they should be maintained per collection.
\item \textbf{Integrated deployment.} The results here are component-level. The
natural next step is an integrated DB$+$LLM benchmark of end-to-end performance and
cost analysis.
\end{itemize}

\section{Conclusion}
Models are increasingly becoming the primary users of our infrastructure. In agentic systems, they already read and write more than humans. 
But they work with token IDs instead of UTF-8 characters. Storing text as the serving model's BPE token IDs gives compression for free, and adding token-ID compression on top surpasses every byte-level algorithm. This also makes both reads and writes from the model significantly faster. 

The bottleneck is mainly standardization. Token-native storage
already pays off today when the store and its model share one tokenizer. The benefits only
grow as the rest of the stack adopts that tokenizer: once embedders, rerankers, and
other model families read the same IDs, one stored copy serves them all, with no
vendor or model-family lock-in. All it takes is a shared vocabulary, ideally in
frequency order, the same kind of agreement that once gave text ASCII and UTF-8.

\end{document}